\documentclass[12pt]{iopart}

\usepackage{iopams}
\usepackage{graphicx}

\begin{document}

\title{Complementary action of chemical and electrical synapses to perception}

\author{F. S. Borges$^1$, E. L. Lameu$^1$, A. M. Batista$^{1,2}$, 
K. C. Iarosz$^3$, M. S. Baptista$^3$, R. L. Viana$^4$}
\address{$^1$ P\'os-Gradua\c c\~ao em Ci\^encias, Universidade Estadual de 
Ponta Grossa, 84030-900, Ponta Grossa, Paran\'a, Brasil.}
\address{$^2$Departamento de Matem\'atica e Estat\'istica, Universidade
Estadual de Ponta Grossa, 84030-900, Ponta Grossa, PR, Brazil.}
\address{$^3$ Institute for Complex Systems and Mathematical Biology,
University of Aberdeen, AB24 3UE, Aberdeen, UK, EU}
\address{$^4$Departamento de F\'isica, Universidade Federal do Paran\'a, 
81531-990, Curitiba, PR, Brazil}
\ead{$^2$antoniomarcosbatista@gmail.com}

\date{\today}

\begin{abstract}
We study the dynamic range of a cellular automaton model for a neuronal network
with electrical and chemical synapses. The neural network is separated 
into two layers, where one layer corresponds to inhibitory, and the other 
corresponds to excitatory neurons. We randomly distribute electrical synapses 
in the network, in order to analyse the effects on the dynamic range. We 
verify that electrical synapses have a complementary effect on the enhancement 
of the dynamic range. The enhancement depend on the proportion of electrical 
synapses, and also depend on the layer where they are distributed.
\end{abstract}

\hspace{4pc}{\small {\it Keywords}: dynamic range, cellular automaton, neuron}

\maketitle

%%%%%%%%%%%%%%%%%%%%%%%%%%%%%%%%%%%%%%%
%%%%%%%%%%%%%%%%%%%%%%%%%%%%%%%%%%%%%%%

\section{Introduction}

The cerebral cortex contains neurons and their fibres \cite{kandel00}. These 
neurons are grouped together into functional or morphological units, called 
cortical areas \cite{rakic88}, each of them playing a well-defined role in the 
processing of information in the brain \cite{essen92}. Hence the theoretical 
understanding of the principles of organisation and functioning of the cerebral
cortex can shed light on the knowledge of many distinct and important subjects 
in neuroscience \cite{buzsaki06}. One relevant subject is psychophysics, that 
analyses the perceptions due to external stimuli \cite{gescheider13}. 

Studies about the relation between sensation and stimulus by measuring the
quantity of both factors were realised by Weber and Fechner \cite{murray93}. 
They proposed that the relation was logarithmic \cite{chialvo06}. However, 
Stevens proposed a theory based on a power-law relation between stimulus and 
response, where the exponent depends on the type of stimulation 
\cite{stevens08}.

The capacity of a biological system to discriminate the intensity of an 
external stimulus is characterized by the dynamic range (DR) \cite{gollo09}. 
DR is a range of intensities for which receptors can encode stimuli 
\cite{stevens08,kinouchi06,batista14}. It is the logarithm of the difference 
between the smallest and the largest stimulus value for which the responses
are not too weak to be distinguished or too close to saturation, respectively.
The lower and upper bounds are arbitrarily chosen due to the fact that the
scaling region is well fit by a power law. In other words, small changes
are not affect our results. The visual and the auditory perception have high 
dynamic range. The human sense of sight can perceive changes in about ten 
decades of luminosity, and the hearing covers twelve decades in a range of 
intensities of sound pressures \cite{chialvo06}. The DR of the human visual is 
important in the design of high dynamic range display devices 
\cite{reinhard10}. Whereas the DR of the hearing is used for cochlear implant 
systems \cite{spahr07}.
 
In this work we study the dynamic range of a cellular automaton with electrical
and chemical synapses \cite{viana14}. We consider that the chemical synapses 
can be excitatory or inhibitory, and a layered model \cite{kurant06},
where one layer consists of excitatory neurons, and the other layer consists
of inhibitory neurons. Network consisting of excitatory and inhibitory neurons 
was considered to describe the primary visual cortex \cite{adini1997}. Pei and 
collaborators \cite{pei12} investigated the behaviour of excitatory-inhibitory 
excitable networks with an external stimuli. They suggested that the dynamic 
range may be enhanced if high inhibitory factors are cut out from the 
inhibitory layer. In our work, we consider a neural network in which neurons 
interact by chemical and electrical synapses in a excitatory-inhibitory
layered model. Our main results are: the equation of the dynamic range for a 
random neural network with chemical and electrical synapses, and to show that 
electrical synapses in the excitatory layer have an influence on the dynamic 
range more significative than in the inhibitory layer, due to the fact that
the electrical synapses in the excitatory layer are responsible for the
complementary effect of dynamic range enhancement. 

This paper is organised as follows: in Section 2 we introduce the cellular
automaton rule, and the random network. Section 3 shows our analytical and
numerical results obtained for the dynamic range. The last section presents the
conclusions.

%%%%%%%%%%%%%%%%%%%%%%%%%%%%%%%%%%%%%%%
%%%%%%%%%%%%%%%%%%%%%%%%%%%%%%%%%%%%%%%

\section{Neuronal network model of spiking neurons}

We consider a cellular automaton model in that a node can spike, $x_i=1$, when 
stimulated in its resting state, $x_i=0$ ($i=1,...,N$). When a spike occurs 
there is a refractory period until the node returns to its resting state, 
$x_i=2,...,\mu-1$. During the refractory period no spikes occur. There are 
excitatory and inhibitory connections linking nodes unidirectionally. The
pre-synaptic node whose out chemical synapses is excitatory (inhibitory)
is called an excitatory (inhibitory) node. Excitatory nodes increase the 
probability of excitation of their connected nodes, while inhibitory nodes 
decrease the probability. The network presents also electrical connections, 
that are bidirectional links. They increase the probability of excitation
between the connected nodes.

\begin{figure}
\begin{center}
\includegraphics[width=24pc,height=14pc]{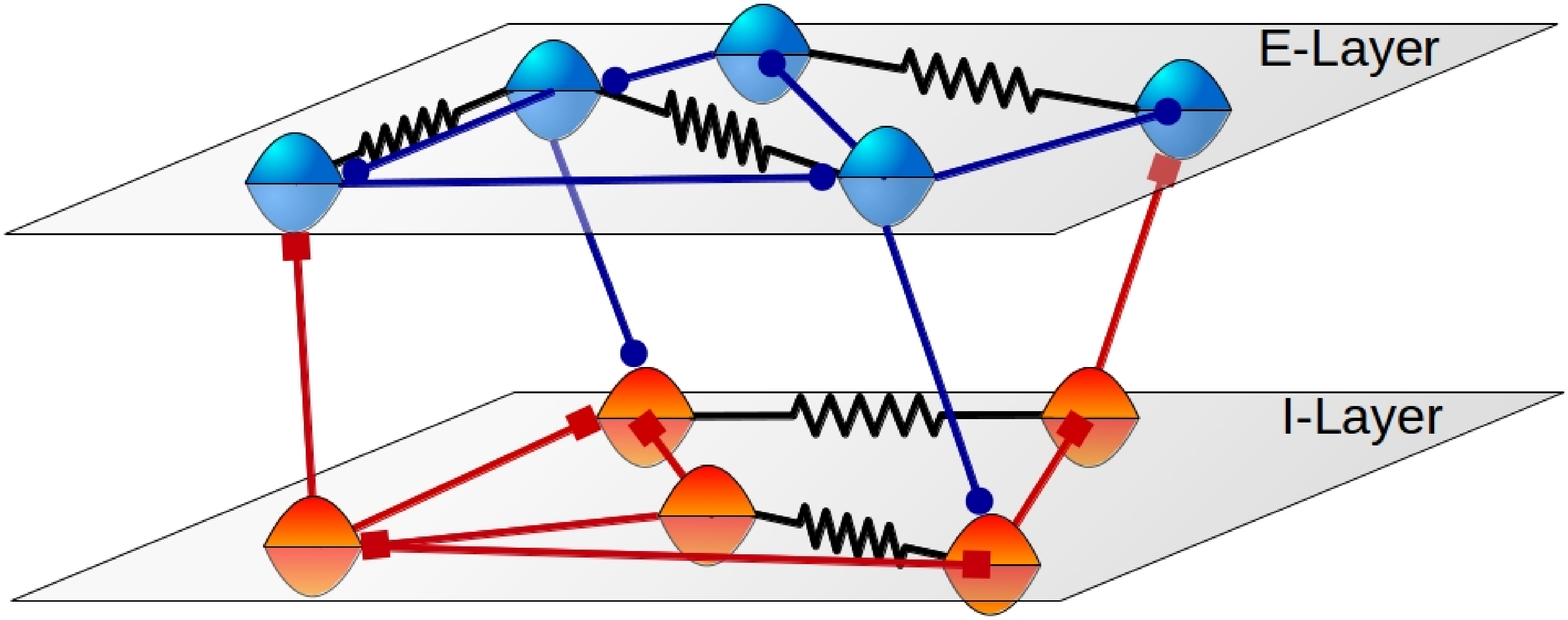}
\caption{Scheme of the E-I layered network with one excitatory layer (E-layer),
and one inhibitory layer (I-layer). The lines with blue filled circles 
represent the excitatory connections, the lines with red filled squares 
represent the inhibitory connections, and the other links represented by black 
saw line are the electrical connections.}
\label{fig1} 
\end{center}
\end{figure}

The dynamics of the cellular automaton is given by: 
\begin{enumerate}
\item 
if $x_i(t)=0$, then 
\begin{itemize}
\item
a node can be inhibited by an excited inhibitory node $j$ ($x_j(t)=1$) with 
probability $B_{ij}$, remaining equal to zero in the next time step; 
\item
a node can be excited by an excited excitatory node $j'$ ($x_{j'}(t)=1$) with
probability $B_{ij'}$;
\item
a node with electrical connection can be excited by an excited node $j''$
with probability $A_{ij''}$.
\item
a node can be excited by an external stimulus with probability $r$.
\end{itemize}
\item
if $x_i\neq 0$, then $x_i(t+1)=x_i(t)+1$ (mod $\mu$), where $x_i(t)\in 
\{0,1,...,\mu-1\}$ is the state of the $i$th node at time $t$. In other words, 
the node spikes ($x_i=1$) and after that remains insensitive during $\mu-2$
time steps.
\end{enumerate}
The weighted adjacency matrices $A_{ij}$ and $B_{ij}$ describe the strength of 
interactions between the nodes. The matrix $A_{ij}$ contains information about
the electrical connections, and the matrix $B_{ij}$ about the excitatory and
inhibitory connections. The connection architecture is described by a random 
graph, in that the connections are randomly chosen \cite{erdos59}. We 
separete the neurons by layers. Figure \ref{fig1} shows the scheme of the 
E-I layered network, where the E layer contains $N_e$ excitatory nodes, and the
I layer contains $N_i$ inhibitory nodes. Then, the layered network has a total 
of $N_e+N_i=N$ nodes. The excitatory connections (blue circles) go from 
excitatory nodes to other nodes, the inhibitory connections (red squares) go 
from inhibitory nodes to other nodes, and the electrical connections are 
bidirectional (black saw lines).

The neuron responses are obtained through the density of spiking neuron
\begin{equation}
p(t)=\frac{1}{N}\sum_{i=1}^{N}\delta(x_i(t),1),
\end{equation}
where $\delta(a,b)$ is the Kronecker delta. With the density we calculate
the average firing rate
\begin{equation}
F=\overline {p(t)}=\frac{1}{T}\sum_{t=1}^{T}p(t),
\end{equation}
$T$ is the time window chosen for the average. 

The update rules for neurons in E and I layer are the same. In order to 
estimate analytically the average firing rate, we calculate the mean-field map 
for $p(t)$ at a long time
\begin{eqnarray}\label{mapp}
p(t+1) & = & [1 - (\mu-1) p(t)] (1- S_{\rm ch} p(t))^{f_{i} K_{\rm ch}} \label{eq2} \\ 
&\times& \{ r + (1-r)[1-(1-S_{\rm ch} p(t))^{f_{e} K_{\rm ch}} 
(1-S_{\rm el} p(t))^{K_{\rm el}}] \},  \nonumber 
\end{eqnarray}
in which $S_{\rm ch}$ is the strength of excitatory and inhibitory interactions, 
$S_{\rm el}$ is the strength of electrical interactions, $f_e=N_e/N$ and
$f_i=N_i/N$ are the fraction of excitatory and inhibitory nodes, respectively.
The average degree of chemical connections is denoted by $K_{\rm ch}$, and 
$K_{\rm el}$ denotes the average degree of electrical connections. The average 
degree is calculated by assuming randomly chosen pairs of nodes.

\begin{figure}
\begin{center}
\includegraphics[width=30pc,height=16pc]{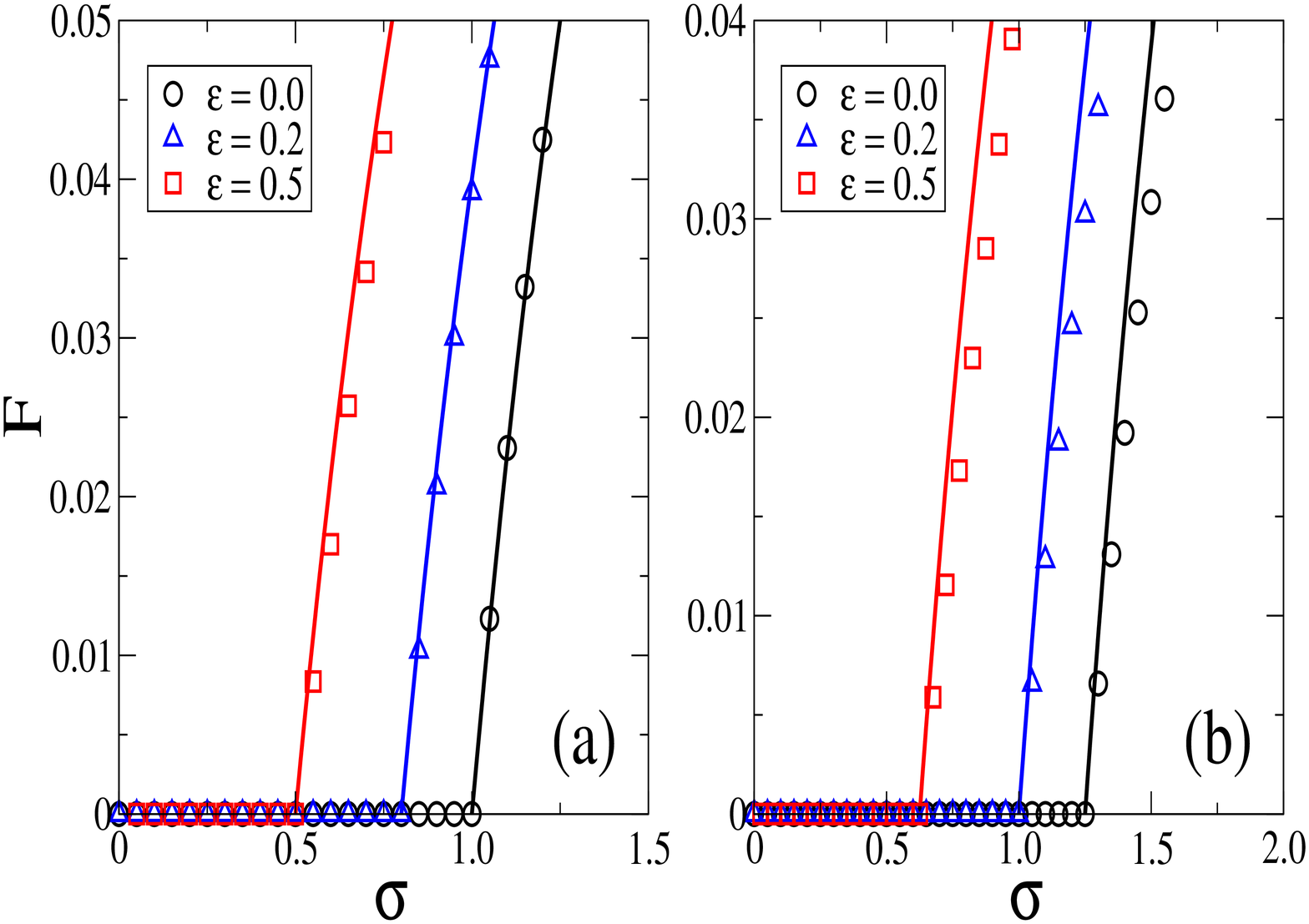}
\caption{(Colour online) Firing rate versus branching ratio $\sigma$ for 
$N=10^5$, $K_{\rm ch}=10$, $S_{\rm el}=1$, $\mu=5$, $r=0$, (a) $f_e=1$, and (b)
$f_e=0.8$. The symbols correspond to simulation results, and the lines are the
theoretical values according to Eq. \ref{papprox}.}
\label{fig2} 
\end{center}
\end{figure}

To obtain an approximate analytical value for the firing rate ($F$) for the 
case of small density of spiking neuron ($p(t)$), and without an external 
perturbation ($r=0$), we expand Eq. (\ref{mapp}) around $p(t)=0$, and find
\begin{equation}\label{papprox}
p^*\approx \frac{\varepsilon + \sigma f_e - 1}{(\mu - 1)(\varepsilon + 
\sigma f_e) + \sigma (\varepsilon + \sigma f_e f_i)}, 
\end{equation}
in a mean-field approximation \cite{kinouchi06} $\sigma=K_{\rm ch}S_{\rm ch}$ is 
the average chemical branching ratio of nodes in the E-layer, and 
$\varepsilon=K_{\rm el}S_{\rm el}$ is the average electrical branching ratio of 
nodes, representing the overall strength of chemical and electrical interaction
in the network. We can write $p^*=p(t+1)=p(t)$ in the stationary state, and as 
a result for large time we have $F\approx p^*$. Figure \ref{fig2} exhibits the 
firing rate varying $\sigma$ for different values of $\varepsilon$. The symbols 
correspond to simulation according to celular automaton rules, and the lines 
correspond to the theoretical values from Eq. (\ref{papprox}).

\begin{figure}
\begin{center}
\includegraphics[width=22pc,height=16pc]{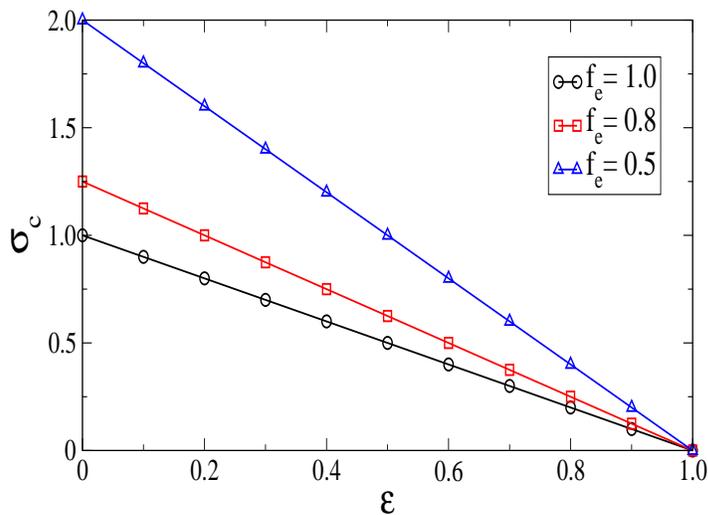}
\caption{(Colour online) $\sigma_c$ versus $\varepsilon$ for $f_e=0.5$ (blue
triangles), $f_e=0.8$ (red squares), and $f_e=1.0$ (black circles). The lines
are from the analytical expression $\sigma_c=(1-\varepsilon)/f_e$.}
\label{fig3} 
\end{center}
\end{figure}

There is a critical value of the average branching ratio $\sigma_c$ in that
the firing rate increases from zero. In other words, $\lim_{r\rightarrow 0}F=0$ if 
$\sigma<\sigma_c$, and $\lim_{r\rightarrow 0}F>0$ if $\sigma>\sigma_c$. We can see 
through Figures \ref{fig2}(a) and (b) that $\sigma_c$ depends on $f_e$ and 
$\varepsilon$. Equation (\ref{papprox}) allows us to obtain the dependence,
given by $\sigma_c=(1-\varepsilon)/f_e$, by assuming that $p$ is null. In 
Figure \ref{fig3} we compare the simulation result (symbols) with the equation 
for $\sigma_c$ (lines). It is possible to see a good agreement. This shows that
chemical and electrical connections complement themselves for obtaining the
critical point $\sigma_c$. The larger (smaller) the electrical branching rate 
in the network, the smaller (larger) the chemical  branching rate must be.

%%%%%%%%%%%%%%%%%%%%%%%%%%%%%%%%%%%%%%%
%%%%%%%%%%%%%%%%%%%%%%%%%%%%%%%%%%%%%%%

\section{Dynamic Range}
 
\begin{figure}
\begin{center}
\includegraphics[width=22pc,height=16pc]{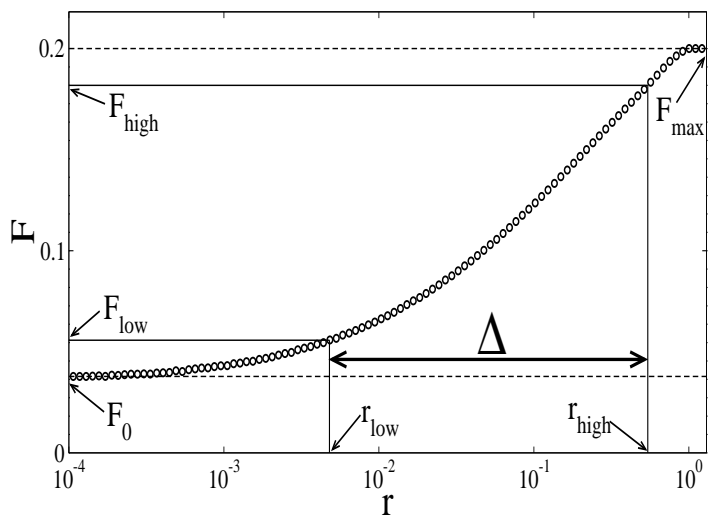}
\caption{Average firing rate as a function of the average input rate.}
\label{fig4} 
\end{center}
\end{figure}

The ratio between the largest and smallest possible values of a changeable 
quantity is called dynamic range. The standard definition of the dynamic 
range is \cite{firestein93}
\begin{equation}
\Delta=10\log_{10}\frac{r_{\rm high}}{r_{\rm low}},
\end{equation}
where $r_{\rm high}$ and $r_{\rm low}$ are the average input rates for 
$F_{\rm high}$ and $F_{\rm low}$, respectively (Fig. \ref{fig4}). The high firing
rate is obtained from $F_{\rm high}=F_0+0.95(F_{\rm max}-F_0)$, and the low firing
rate is from $F_{\rm low}=F_0+0.05(F_{\rm max}-F_0)$, where $F_0$ is the value for
the minimum saturation, and $F_{\rm max}$ is the value for the maximum 
saturation. If the system has a refractory time $\mu$, the maximum firing rate 
$F_{\rm max}$ is equal to $1/\mu$.

Taking the limit of Eq. (\ref{mapp}) as $p$ approaches zero
\begin{equation}
p\approx [1-(\mu-1)p]{\rm e}^{-f_i\sigma p} 
[r+(1-r)(1-{\rm e}^{-p(f_e\sigma+\varepsilon)})],
\end{equation}
which allow us to obtain $r_{\rm low}$ by doing $r=r_{\rm low}$. Then, the dynamic 
range is given by
\begin{eqnarray}\label{eqDR}
\Delta=-10\log_{10} \left[\frac{1-{\rm e}^{F_{\rm low} (f_e \sigma + \varepsilon)}}
{r_{\rm high}}+ \frac{F_{\rm low} {\rm e}^{F_{\rm low} (\sigma +\varepsilon )}}{r_{\rm high}
-(\mu-1) F_{\rm low} r_{\rm high}} \right]. 
\end{eqnarray}
We have verified that $r_{\rm high}$ is approximately equal to $0.75$ for our
simulation considering $N=10^5$, and $\mu=5$. By means of $F_0$ (Eq. 
\ref{papprox}) the firing rate $F_{\rm low}$ can be calculated through relation 
$F_{\rm low}=F_0+0.05(F_{\rm max}-F_0)$.

\begin{figure}
\begin{center}
\includegraphics[width=34pc,height=16pc]{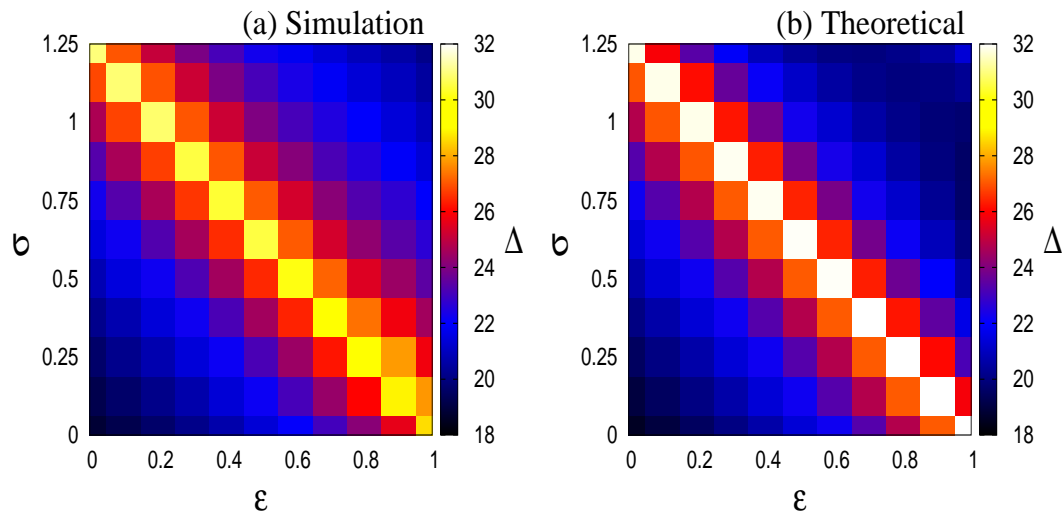}
\caption{(Colour online) Dynamic range as a function of $\sigma$ and 
$\varepsilon$ for $N=10^5$, $K_{\rm ch}=10$, $S_{\rm el}=1$, $f_e=0.8$, and 
$\mu=5$. The simulation is showed in (a), and the result according to Eq.
(\ref{eqDR}) is showed in (b).}
\label{fig5} 
\end{center}
\end{figure}

In Figure \ref{fig5} we compare the dynamic range calculated using simulation 
(a) and from Eq. (\ref{eqDR}) (b). We can see that the dynamic range
is maximum for $\sigma_c$, which is obtained by assuming that $r=0$. In the 
subcritical region for $\sigma f_e+\varepsilon<1$ it is possible to observe 
that weak stimuli are amplified and the sensitivity is enlarged, as a result of
the activity propagation among neighbours. Therefore, the dynamic range 
increases with $\sigma$ and $\varepsilon$. In the supercritical region for 
$\sigma f_e+\varepsilon>1$ the dynamic range decreases due to the fact that
the average firing rate is positive, and masks the effect of weak stimuli. 
The theoretical result shows a good agreement with the simulation, except for 
the supercritical region. This occurs due to the fact that the values of $p(t)$
are not around zero in the supercritical region, and we have considered $p(t)$ 
around zero to obtain analytical results.

\subsection{Influence of electrical synapses}

Pei and collaborators \cite{pei12} investigated a excitatory-inhibitory
excitable cellular automaton considering undirected random links. They verified
that the dynamic range can be enhanced if the nodes with high inhibitory factors
in the inhibitory layer are cut out. In this work, we are considering not only
chemical synapses (directed links), but also electrical synapses (undirected 
links). Our interest is to understand the role of the electrical synapse in
the dynamic range. For that goal Figure \ref{fig5} shows the dynamic range for
a network that has electrical synapses randomly distributed in all the 
network. We make further analysis considering the effect of electrical synapses 
on the dynamic range in two cases: (i) randomly distributed in the excitatory 
layer, and (ii) randomly distributed in the inhibitory layer.

\begin{figure}
\begin{center}
\includegraphics[width=34pc,height=16pc]{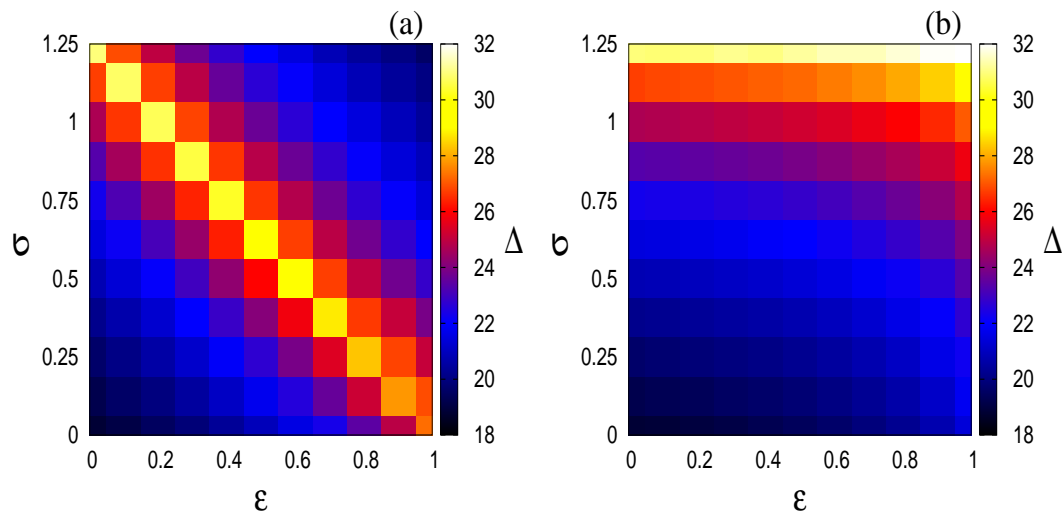}
\caption{(Colour online) Dynamic range as a function of $\sigma$ and 
$\varepsilon$ for $N=10^5$, $K_{\rm ch}=10$, $S_{\rm el}=1$, $f_e=0.8$, and 
$\mu=5$. We consider randomly distributed electrical synapses in (a) excitatory
layer, and (b) inhibitory layer.}
\label{fig6} 
\end{center}
\end{figure}

Figure \ref{fig6}(a) shows the dynamic range for a network where electrical 
synapses are distributed in the excitatory layer. We can see that the behaviour
of the dynamic range is similar to the one reported in Figure \ref{fig5}. The 
maximum dynamic range follows the equation for $\sigma_c$, namely the dynamic 
range can be enhanced increasing the amount of electrical synapses with the 
decrease of chemical synapses in the excitatory layer. However, when the 
electrical synapses are randomly distributed in the inhibitory layer, we do not
observe a similar behaviour for the maximum dynamic range. In this case the 
dynamic range is affected by electrical synapses, but it is mainly enhanced by 
chemical synapses.

%%%%%%%%%%%%%%%%%%%%%%%%%%%%%%%%%%%%%%%
%%%%%%%%%%%%%%%%%%%%%%%%%%%%%%%%%%%%%%%

\section{Conclusions}

We have modelled a neuronal network using cellular automaton which models the
behaviour of a neural network where neurons interact by electrical and chemical
synapses. The chemical synapses are separated into two layers, where one layer 
corresponds to excitatory, and the other corresponds to inhibitory.

Our aim has been to determine the dynamic range as a function of the types
os synapses. Reference \cite{pei12} shows theoretical analysis and simulations 
considering undirected connections. In this work, we have considered undirected
electrical and directed chemical connections. Electrical and chemical synapses 
are relevant in cells found in the retina \cite{hidaki05,publio12}, and cells 
in olfactory bulb \cite{kosaka05}.

We have obtained theoretical results for the average firing rate, and for the
critical value of the average branching ratio that exits the network, allowing
it to fire and consequently allowing information to be transmitted. From the
equation of the average firing rate we obtained an equation for the dynamic 
range. This equation shows that the dynamic range is maximum for the critical
average branching ratio. The equation presents a remarkable agreement with 
our simulations, mainly around the critical average branching rate. 
We verified an increase of the dynamic range in the subcritical region, and
a decrease in the supercritical region. As a result, we verified that the 
enhacement of the dynamic range depends on the parameters $\varepsilon$ and
$\sigma$ that are associated with the electrical and chemical synapses.
Moreover, our results show that electrical synapses in the excitatory layer 
have an influence on the dynamic range more significative than when electrical 
synapses are placed in the inhibitory layer. The complementary effect occurs
due to electrical synapses in the excitatory layer.

%%%%%%%%%%%%%%%%%%%%%%%%%%%%%%%%%%%%%%%
%%%%%%%%%%%%%%%%%%%%%%%%%%%%%%%%%%%%%%%

\ack
This study was possible by partial financial support from the following 
Brazilian government agencies: Funda\-\c c\~ao Arauc\'aria, EPSRC-EP/I032606/1,
CNPq, CAPES and Science Without Borders Program - Process n$^o$ 17656125,
n$^o$ 99999.010583/2013-00 and n$^o$ 245377/2012-3.

%%%%%%%%%%%%%%%%%%%%%%%%%%%%%%%%%%%%%%%
%%%%%%%%%%%%%%%%%%%%%%%%%%%%%%%%%%%%%%%

\end{document}